\documentclass[journal]{IEEEtran}
\usepackage{cite}

\ifCLASSINFOpdf
\else
\fi
\usepackage{amsmath}
\usepackage[ruled,vlined,linesnumbered]{algorithm2e}

\usepackage{array}
\usepackage{fixltx2e}
\usepackage{url}

\usepackage[colorinlistoftodos,prependcaption,textsize=tiny]{todonotes}
\usepackage{xcolor}
\usepackage[normalem]{ulem}

\usepackage{tikz}
\usepackage{pgfplots}

\begin{document}
	%
	\title{Statistical Clear Sky Fitting Algorithm}
	%
	%
	%
	
	\author{Bennet~Meyers,~\IEEEmembership{Member,~IEEE,}
		Michaelangelo~Tabone,~and~Emre~Can~Kara
		\thanks{B. Meyers is with the Department of Electrical Engineering at Stanford University e-mail: (bennetm@stanford.edu).}
		\thanks{M. Tabone is with the Department of Civil and Environmental Engineering at Stanford University; E. C. Kara is with the Grid Integration, Systems and Mobility group at SLAC National Accelerator Laboratory.}
		\thanks{Manuscript submitted May 21, 2018}}
	
	%
	%

	\markboth{}%
	{Shell \MakeLowercase{\textit{et al.}}: Statistical Clear Sky Fitting Algorithm}
	%



	\maketitle
	
	\begin{abstract}
		We present an algorithm that estimates a clear sky performance signal from the measured power of a PV system. The algorithm uses only observed power output, and assumes no knowledge of weather, irradiance data, or system configuration metadata. This is a novel approach to understanding the clear sky behavior of an installed PV system, that does not rely on traditional atmospheric and geometric modeling techniques.
		%
		%
	\end{abstract}
	

	%
	\IEEEpeerreviewmaketitle

	\section{Introduction}
	%
	%
	%
	%
	\IEEEPARstart{A} clear sky condition is generally defined as the absence of visible clouds in a given location. A more rigorous definition of a clear sky condition involves capturing the effects of precipitable water, aerosols, ozone, and other atmospheric phenomena on irradiance transmittance \cite{Younes2007}. However, as discussed in \cite{Reno2016}, ``clear sky equivalent conditions'' can be thought of as local atmospheric conditions that create an irradiance profile that is approximately equal the irradiance observed during true clear sky conditions.  Approximate clear sky irradiance is a useful concept for assessing photovoltaic (PV) system performance, for forecasting the performance of future systems, and possibly for identifying the sources of system losses \cite{Reno2016}. 
	
	Traditional clear sky models provide estimates of global horizontal irradiance (GHI), direct normal irradiance (DNI), and diffuse horizontal irradiance (DHI), by modeling how the atmosphere attenuates extra-atmospheric irradiance under non-cloudy conditions. Clear sky models are often transposed to estimate the plane of array irradiance (POAi) on a PV system. From here, an analyst may use the modeled clear sky POAi to estimate the power output of a system under clear conditions, or use the modeled value as a reference to identify approximately clear time periods in a historic dataset \cite{Reno2016}. A review of common clear sky models is given in \cite{Ineichen2016}; these models are implemented in Python as part of the \texttt{pvlib-python} software package \cite{Holmgren2015}.
	
	In this paper, we consider a PV system's approximate clear sky signal to be a \textit{baseline} for the observed power signal. Deviations from the baseline primarily represent the effects of atmospheric phenomenon but could also include degradation modes such as soiling and operational issues such as inverter failure. Qualitatively, we are looking for a signal that is smooth over the course of a day, changes slowly from day-to-day (approximate daily periodicity), and repeats on a yearly timescale.  These qualitative characteristics can be captured using statistical signal processing and optimization techniques to estimate the clear sky signal from observed power measurements.
	
	This work proposes to invert the clear sky estimation process by calculating it directly from historical data without the need to simulate solar geometry or to estimate/observe atmospheric conditions. Our proposed method is based on generalized low-rank modeling \cite{Udell2014} for determining the baseline of a stochastic process that exhibits \textit{cyclostationarity}, i.e.~a signal that has statistical properties that vary cyclically with time. The framework requires the selection of regularization functions and associated hyperparameters that depend on the specific application. We implement this framework in the form of a statistical clear sky fitting (SCSF) algorithm to estimate clear sky signal from the observed power generation of a PV system.
	
	Notably, the SCSF algorithm requires no data besides the raw power production signal. As a point of comparison, the Ineichen clear sky model takes as inputs the extraterrestrial irradiance, solar zenith angle, air mass, Linke turbidity factor, and elevation \cite{Ineichen2002}, while the Simplified Solis model parameterizes clear sky irradiance in terms of precipitable water and aerosol optical depth \cite{Ineichen2008}. In addition, the SCSF algorithm is applicable to a wide range of time frames (weeks to years) and is robust to missing or incorrect data. Thus, it can be applied to systems that lack reliable locational and mounting metadata, or in locations where the reference data for Linke turbidity factors, precipitable water, and aerosol optical depth are unknown or not trusted. The algorithm relies on mathematical theory from linear algebra, signal processing, and convex optimization to exploit daily and yearly periodic structure in the time-series power data.
	
	\section{Methodology}
	\subsection{Background}
	Consider the discrete signal $\left\{x[t] : t=1, 2,\dots\right\}$, sampled at uniform intervals. We can consider the observation of $T$ consecutive elements of this signal to be a vector $x\in\mathbf{R}^T$. Suppose that this signal exhibits approximate periodic structure on a period $m$. In other words, $x_i\approx x_{i+m}$ for $i=1,\cdots, (T-m)$ (possibly in the presence of significant noise). Without loss of generality, we can assume that $T$ is a multiple of $m$ such that $T=m\cdot n$. Then, we can consider forming a matrix $M\in\mathbf{R}^{m\times n}$ where
	\begin{align*}
	M = \left[\begin{matrix}
	x_1 & x_{m+1} & \cdots & x_{(n-1)m+1}\\
	x_2 & x_{m+2} & \cdots & x_{(n-1)m+2} \\
	\vdots & \vdots	&\ddots & \vdots \\
	x_m & x_{2m} & \cdots &x_{nm}
	\end{matrix}\right].
	\end{align*}
	Generalized low-rank modeling \cite{Udell2014} is concerned with finding a matrix $Z$ that solves the optimization problem
	\begin{alignat*}{2}
	& \underset{Z}{\text{minimize}}
	& & \quad \left\lVert M - Z \right\rVert_F\\
	& \text{subject to} 
	& & \quad \text{Rank}(Z)\leq k.
	\end{alignat*}
	Typically, this rank constraint is encoded by defining $Z$ as the product of two low-rank matrices, $L\in\mathbf{R}^{m\times k}$ and $R\in\mathbf{R}^{k\times n}$. While the factorization is not unique, there are many approaches to solving this problem. The Eckart-Young Theorem \cite{Eckart1936} states that a solution to this problem is given by truncating the Singular Value Decomposition (SVD) of $M$ to the top $k$ singular values. However, we are interested in the more general regularized low-rank model, in which additional constraints are put on $L$ and $R$:
	\begin{alignat*}{2}
	& \underset{L\in \mathbf{R}^{m\times k},R\in \mathbf{R}^{k\times n}}{\text{minimize w.r.t $\mathbf{R}^{3}$}}
	& & \quad \left(\left\lVert M - LR \right\rVert_F, \phi_1(L), \phi_2(R) \right),
	\end{alignat*}
	where $\phi_1:\mathbf{R}^{m\times k}\rightarrow \mathbf{R}$ and $\phi_2:\mathbf{R}^{k\times n}\rightarrow \mathbf{R}$ are regularization functions that act on the component matrices. This is a multicriterion optimization problem, in which we weight the regularization terms via the parameters $\mu_1$ and $\mu_2$:
	\begin{alignat*}{2}
	& \underset{L\in \mathbf{R}^{m\times k},R\in \mathbf{R}^{k\times n}}{\text{minimize}}
	& & \quad \left\lVert M - LR \right\rVert_F +\mu_1 \phi_1(L) +\mu_2 \phi_2(R) ,
	\end{alignat*}
	When both regularization functions are the Frobenius norm $\left(\lVert \cdot \rVert_F\right)$, this problem has a closed-form solution \cite{Udell2014}. However, when these are more general penalty or loss functions, this objective may be minimized through alternating minimization---simply alternating between minimizing over $L$ keeping $R$ fixed, and then  minimizing $R$ keeping $L$ fixed. If $\phi_1$ and $\phi_2$ are convex functions, then each step in this process involves solving a convex minimization problem. Using this technique to estimate the baseline signal for a cyclostationary physical process (such as the power output of a PV system over time) involves the careful selection of regularization functions and hyperparameters that are specific to the given process.
	
\begin{figure}
		\centering
		\includegraphics[width=9cm]{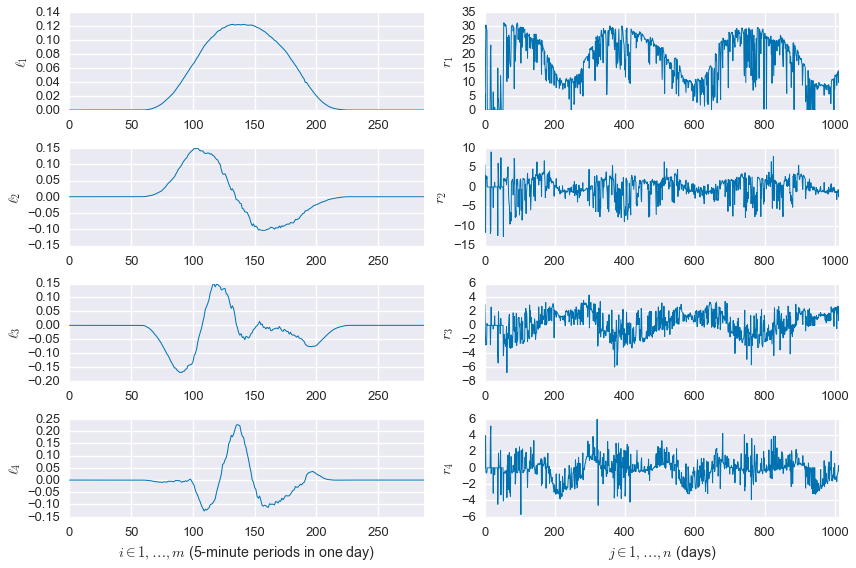}
		\caption{The top $4$ columns of $L$, $\left(\ell_1, \ell_2, \ell_3, \ell_4\right)\in\mathbf{R}^m\times \mathbf{R}^4$, and rows of  $R$, $\left(r_1, r_2, r_3, r_4\right)\in\mathbf{R}^n\times \mathbf{R}^4$, when SVD is used to factor $M$. Note the wavelet-like shape of the left vectors and the strong seasonality of the right vectors.}
		\label{fig:svd_decomp}
	\end{figure}

	This approach is well-suited to estimate the baseline signal for a cyclostationary physical process (such as the power output of a PV system over time) because it decomposes the time-series signal into a low-rank representation of variations within a single period of the signal and a low-rank representation of how the period changes over time. Some intuition for this behavior and why low rank models can be useful for analyzing PV data can be gained by inspecting the SVD for a typical PV power signal, as shown in Figure (\ref{fig:svd_decomp}). (Recall that SVD gives an optimal low-rank factorization in the absence of regularization.) By selecting appropriate regularization terms, it is possible to select for a factorization that isolates the clear sky signal in the data, as described in the following section.
	
	%
	%
	\subsection{The SCSF Problem} \label{scsf-problem}
	Let $D\in\mathbf{R}^{m\times n}$ be a matrix containing the measured power signal from the PV system, segmented into daily chunks, where $m$ is the number of data sampled in one day (24 for 1-hour data, 288 for 5-minute data) and $n$ is the number of days in the data set. Then we introduce the following data model:
	\begin{align}
	D \approx LR,
	\end{align}
	where $L\in\mathbf{R}^{m\times k}$ and $R\in\mathbf{R}^{k\times n}$. We would like to find low-rank matrices $L$ and $R$ which estimate the output of the system under approximately clear sky conditions. We introduce the qualitative characteristics we want the clear sky signal to have:
	\begin{enumerate}
		\item The signal should be smooth over the course of any single day.
		\item The signal should change slowly from day to day.
		\item The signal should closely match actual clear days in the data set.
		\item Negative deviations from the signal are more likely than positive deviations.
		\item The signal is 365-day periodic, with an offset due to long-term degradation
	\end{enumerate}
	These characteristics guide the design of the regularization functions we will use. In addition, we alter the primary objective function $\left\lVert D- LR\right\rVert_F$ to encode the last two conditions above. To find such matrices $L$ and $R$, we pose the following optimization problem:
	\begin{alignat}{2}
	& \underset{L,R}{\text{minimize}}
	& & \quad \sum_{i=1}^{4}f_i \label{prob-main}\\
	& \text{subject to} 
	& & \quad LR \geq 0 \label{constraint-1}\\
	&&&\quad \mathbf{1}^T\ell_j=0,\quad j=2,\ldots, k  \label{constraint-2}\\
	&&&\quad L_{i,j} = 0,\quad i\in\mathcal{Z},\; j=1,\ldots, k  \label{constraint-3}
	\end{alignat}
	where $\ell_j$ is the $j^{\text{th}}$ column of $L$ and $\mathcal{Z}=\left\{i:\left[D\mathbf{1}\right]_i \leq \epsilon \right\}$ is a set defining the indices of times during which the sun is down over the entire year (i.e. the index corresponding to 1:00 AM is in the set. Whereas, the index corresponding to 6:00 AM is not in the set because the sun might have risen by that time during a portion of the year). The functions forming the objective are defined as
	\begin{align}
	f_1 &= \phi_{\tau}\left( \left(D - LR\right) \mathbf{diag}(w)\right) \label{eqn-f1}\\
	f_2 &= \mu_L \left\lVert \mathcal{D}_2 L \right\rVert_F \\
	f_3 &= \mu_R \left\lVert \mathcal{D}_2 R^T \right\rVert_F \\
	f_4 &= \begin{cases}
	\mu_R \left\lVert \mathcal{D}_{1,365} \tilde{R}^T\right\rVert_F &\text{if }n > 365 \\
	0&\text{otherwise}\end{cases}
	\end{align}
	
	$L\in\mathbf{R}^{m\times k}$ and $R\in\mathbf{R}^{k\times n}$ are the decision variables for this problem, and $\phi_{\tau}$ is the \textit{tilted $\ell_1$ penalty} (see section \ref{opt-description}). The functions themselves and their relationship to the five qualitative characteristics of a clear sky signal are explained in detail in the following subsection. The symbol $\mathbf{1}$ represents a vector with every entry equal to $1$ and can be thought of as an element-wise summation operator for a vector.   $\mathcal{D}$ is being used to represent a set of ``difference'' matrices. The simplest is $\mathcal{D}_1$, which has all $-1$'s on the diagonal and all $1$'s on the first superdiagonal. $\mathcal{D}_1$ takes the first order difference of its input vector, returning a vector with length one less than the input $\left(y=\mathcal{D}_1x\implies y_k=x_{k+1} - x_k\right)$. $\mathcal{D}_2$ takes the second order difference of its input vector, retuning a vector with length two less than the input. In other words, if
	$$y = \mathcal{D}_2x\in\mathbf{R}^{n-2} $$
	then,
	$$y_k = x_k - 2x_{k+1} + x_{k+2}$$
	The matrix $D_{1,365}$ takes the first order difference with an offset of 365; it subtracts pairs of elements 365 locations apart. Because all of these operations can be expressed in matrix form, they are linear functions of the input variable.
	
	\subsection{Description of optimization problem components} \label{opt-description}
	\textit{Constraints:} Problem (\ref{prob-main}) includes three hard constraints on the matrices $L$ and $R$. The first, Eqn. (\ref{constraint-1}), constrains the estimated clear sky power to be non-negative (a physical constraint). The second, Eqn. (\ref{constraint-2}), constrains all columns of $L$ except for the first to sum to zero. This constraint ensures that the daily energy production is fully contained in the first column of $L$. The subsequent columns of $L$ correct the shape of the daily power signal, without changing the overall energy. The third constraint, Eqn.  (\ref{constraint-3}), limits the number of free variables in the problem by setting nighttime values to be exactly zero. 
	%
	%
	
	\textit{Function 1:} The first function minimizes the residuals between the model and the given data with respect to the tilted $\ell_1$-penalty \cite{BoydShortCourse}. For $\tau\in(0,1)$, this penalty is defined as
	\begin{align*}
	\phi_{\tau}(x) = \tau(x)_{+} + (1-\tau)(u)_{-}=\frac{1}{2}\left\lvert x \right\rvert + \left(\tau-\frac{1}{2}\right)x
	\end{align*}
	
	In general, this cost function causes a portion, $\tau$, of the $x$'s to be less than or equal to zero, while encouraging many of the $x$'s to be exactly zero. In this sense, it a heuristic for sparseness much in the same way as the standard $\ell_1$-norm, while allowing for unbalanced residuals. In this application, we set $\tau=0.9$, which strongly encourages negative residuals, capturing the fourth qualitative characteristic described in the previous subsection.
	
	$L$ and $R$ provide a low-rank approximation of the given data, with intra-day variability described by $L$ and inter-day variability described by $R$. We will select matrices that produce smooth daily signals that change slowly over time, i.e. they are approximately periodic on daily and yearly time scales. $\hat{D}_{\text{cs}}=LR$ is the clear sky estimate of $D$. The vector $0\preceq w\preceq\mathbf{1}$ is a vector of daily weights, calculated from the problem data. These weights are a rough heuristic for finding individual clear days, by giving large weights to smooth days that have high energy production (compared to the local average). Days with energy much lower than the local average and days that have very rough power signals are given zero weights, excluding them completely from the fitting procedure, satisfying characteristic 3.
	
	\textit{Function 2:} This function penalizes the second-order difference along the columns of $L$, enforcing smoothness in intra-day variability. In other words, this results in a smooth clear sky estimate (characteristic 1).
	
	\textit{Function 3:} This function penalizes the second-order difference along the rows of $R$, enforcing smoothness in inter-day variability. This ensures that the daily clear sky signal changes slowly from day to day, capturing the approximate daily periodicity of the signal (characteristic 2).
	
	\textit{Function 4:} This function penalizes the 365-day lagged first-order difference along the rows of $\tilde{R}\in\mathbf{R}^{(k-1)\times n}$, which is $R$ with the first row removed. This enforces the approximate yearly periodicity of the signal (characteristic 6). This function is only active when more than one year of data is passed to the algorithm. In the case that the function is active, we also include the following additional constraints
	\begin{align*}
	R_{1,j} & = R_{1,j+365} + \beta, \quad j=1,\ldots, m,
	\end{align*}
	where $\beta\in\mathbf{R_+}$ is a new decision variable for the problem. This linear constraint relaxes the requirement on the entries of the first row of $R$ to be 365-day periodic. As mentioned in the description of the constraints, the first column of $L$ captures the energy content of the signal. The entries of the first row of $R$, therefore, correlate exactly with total daily energy. So, this new constraint allows the year-over-year difference in energy to differ by the same set amount over all possible pairs of days. The variable $\beta$ in turn represents the overall degradation rate of the system. 
	
	\subsection{Presentation of Algorithm}
	
	\begin{algorithm}[t!]
		\label{alg-scsf}
		\caption{Statistical Clear Sky Fitting (SCSF)}
		\KwIn{$\left\{p[t]:t=1,\ldots, T\right\}$, the time-series power signal produced by a PV system}
		\KwOut{$\left\{\hat{p}_{\text{cs}}[t]:t=1,\ldots, T\right\}$, the estimated clear sky signal associated with the input data}
		Form matrix $D\in\mathbf{R}^{m\times n}$ from signal $p[t]$ with each day of data arranged in a separate column and $T = m\times n$. \\
		Take singular value decomposition (SVD): $D=U\Sigma V^T$.\\
		Select top $k$ singular values (typically 5-10)\\
		Initialize $L=U$ and $R=\Sigma V^T$ \\
		Calculate weight vector $w$\\
		\Repeat{convergence}{
			Solve prob.~(\ref{prob-main}) in terms of $L$, holding $R$ constant\\
			Solve prob.~(\ref{prob-main}) in terms of $R$, holding $L$ constant
		}
		Form $\hat{p}_{\text{cs}}[t]$ by flattening $LR$ using column-major order \\
		\Return{$\hat{p}_{\text{cs}}[t]$}
	\end{algorithm}
	
	The SCSF algorithm takes an observed time-series power signal as an input and returns the estimated clear sky power output of the system over the course of a year. This is achieved by solving the non-convex optimization problem posed in (\ref{prob-main}). This problem may be solved through the method of alternative minimization decribed in \cite{Udell2014}, extended to include the additional variable, $C$. Because this approach does not guarantee finding the global minimum, it is sensitive to initialization points. A reasonable initialization is a factorization based on SVD for $L$ and $R$ as follows: If $D = U\Sigma V^T$, then select $L$ to be the first $k$ columns of $U$ and $R$ to be the first $k$ rows of $\Sigma V^T$.
	
	The objective is nonincreasing at every iteration and therefore bounded. It is, however, theoretically possible for the alternating minimization method to fail. For example, the function may have stationary points that are not solutions of problem (\ref{prob-main}). However, in practice and with well-conditioned hyperparameters, this approach yields a useful solution to the posed problem.
	
	Convergence of the algorithm can be reasonably taken to be either when the change in the objective function is below a threshold or when a maximum number of iterations are reached. For all data sets explored in this paper, the objective function changed by less than $0.1\%$ after about 10 iterations of the algorithm, which was taken as the convergence criterion.
	
	\section{Results and Validation}\label{section-results}
	In this section, we evaluate the performance and usefulness of the algorithm. First, we show that the algorithm produces a reasonable estimate of clear sky power for a variety of input signals. Then we provide a validation that the algorithm is not overfit to the data through a residual analysis. Finally, we show that the algorithm is perfectly capable of recreating a synthetic clear sky signal, as generated by the \texttt{pvlib-python} modeling chain.
	\subsection{Visual verification}
	\begin{table}[!t]
		\caption{Summary of Validation Datasets}
		\label{table-systemsummary}
		\centering
		\begin{tabular}{|c|cccc|}
			\hline
			& Location & Time Span & Resolution & Size \\
			\hline 
			System 1 & Orange County, CA & 6 months & 5-minute & 482\\ 
			System 2 & Orange County, CA & 3 years & 5-minute & 1408\\
			System 3 & San Mateo County, CA & 1 year & 1-minute & 1815\\
			\hline
		\end{tabular}
	\end{table}
	\begin{figure}[t!]
		\centering
		\includegraphics[width=9cm]{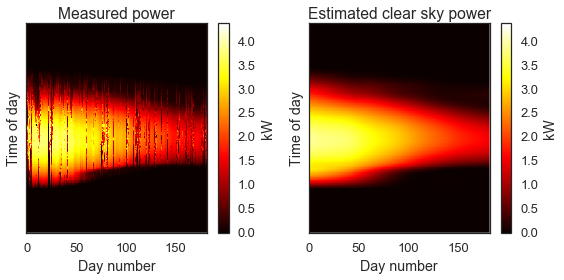}
		\caption{The measured power and clear sky power estimated by SCSF for system 1.}
		\label{fig:sys1_heatmap}
	\end{figure}
	
	\begin{figure}[t!]
		\centering
		\includegraphics[width=9cm]{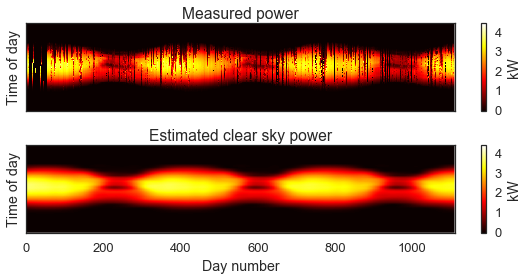}
		\caption{The measured power and clear sky power estimated by SCSF for system 2. This system experiences significant mid-day shade in the winter, which is included in the SCSF estimate as it is truly part of the clear sky signal for this system.}
		\label{fig:sys2_heatmap}
	\end{figure}
	
	\begin{figure}[t!]
		\centering
		\includegraphics[width=9cm]{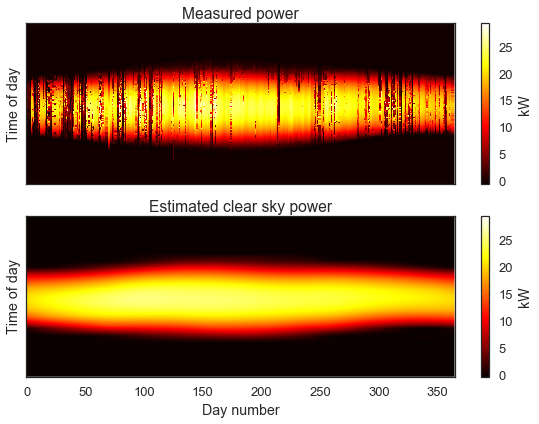}
		\caption{The measured power and clear sky power estimated by SCSF for system 3.}
		\label{fig:sys3_heatmap}
	\end{figure}
	We apply the algorithm to three datasets consisting of measured AC power at three different sites, described in Table~\ref{table-systemsummary}. The size column in this table lists the number of decision variables in each problem instance. Figures \ref{fig:sys1_heatmap}--\ref{fig:sys3_heatmap} illustrate the measured power and estimated clear sky power for each of the three systems as a heatmap. Note that the algorithm preserves the seasonal behavior of the underlying data, including hard site shading, which is especially notable in Figure \ref{fig:sys2_heatmap}. These visualizations suggest that the algorithm can be interpreted as a type of filter. They also enforce the periodic structure of the underlying data, on which the algorithm is built.
	
\begin{figure}[t!]
		\centering
		\includegraphics[width=8cm]{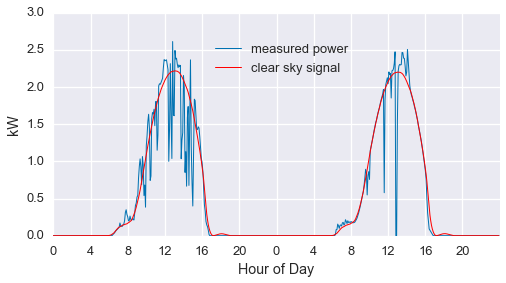}
		\caption{The measured power and clear sky power estimated by SCSF for system 1 over two days. This system is experiencing partial morning shade and near-complete afternoon shade, resulting in a highly irregular daily curve shape that would not be well-fit by a classic modeling pipeline.}
		\label{fig:sys1_ts}
	\end{figure}

\begin{figure}[t!]
		\centering
		\includegraphics[width=8cm]{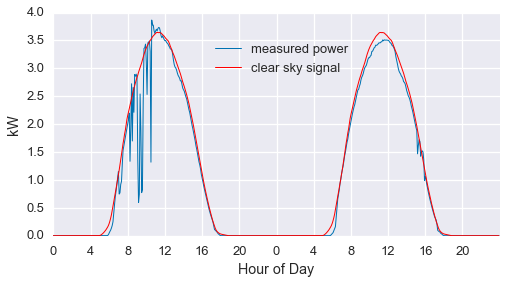}
		\caption{The measured power and clear sky power estimated by SCSF for system 2 over two days.}
		\label{fig:sys2_ts}
	\end{figure}

\begin{figure}[t!]
		\centering
		\includegraphics[width=8cm]{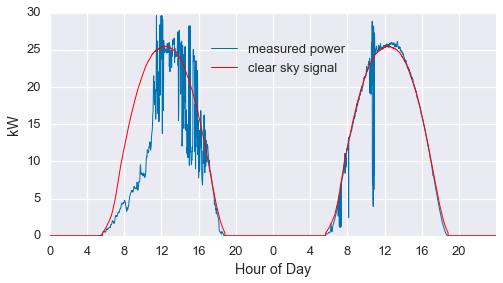}
		\caption{The measured power and clear sky power estimated by SCSF for system 3 over two days.}
		\label{fig:sys3_ts}
	\end{figure}

	Figures \ref{fig:sys1_ts}--\ref{fig:sys3_ts} show the measured power and estimated clear sky power for two days of the given data set. None of the days illustrated in these figures were used to fit the model, as they were excluded by the weighting scheme. So, the agreement between the model and the clear periods in the plotted days are a strong validation of the SCSF algorithm.
	
	\subsection{Residual analysis / test for overfitting}
	The SCSF algorithm can be viewed as a statistical model training method. In this context, it is important to verify that that the model we have developed is \textit{generalizable}, i.e.~it performs well on data that the model was not trained on. In some sense, this validation is built into the algorithm. As previously mentioned, the algorithm only uses a small number of days in the dataset which have non-zero weights. So, the very ability of the algorithm to accurately predict the clear sky power on cloudy days in the data set, as shown in the previous subsection, illustrates that the model is not overfit to the clear days in the data set.
	
\begin{figure}[t]
		\centering
		\includegraphics[width=4cm]{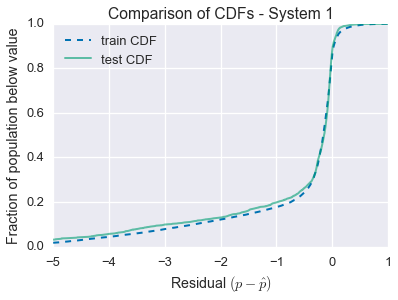}
		\includegraphics[width=4cm]{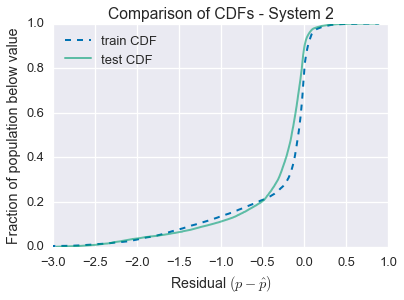}
		\includegraphics[width=4cm]{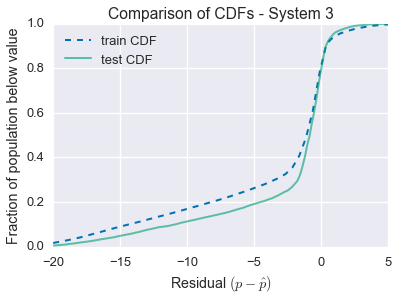}
		\caption{A comparison of train and test residual CDFs for the three datasets. The KS test statistics for these three comparisons change depending on the randomly selected days to be in the reserved test set, but some representative numbers are $0.035$, $0.182$, and $0.095$  respectively. For these translate to $\alpha$~
			thresholds of less than $1\text{e-}10$. In other words, it is very unlikely that these residuals were drawn from different distributions.}
		\label{fig:cdfs}
	\end{figure}

	We make this assessment more rigorous through the following procedure. First, a subset of the problem data is set aside (the ``test set''). We randomly select $10\%$ of the available days for each set of data. Then, the model is trained on the rest of the data (the ``training set''). After the clear sky signal is estimated, the residuals $\left(p_{\text{measured}} - \hat{p}_{\text{clear sky}} \right)$ are calculated for the train and test sets. If the distribution of the residuals from the test set match that of the train set, then the model is not overfit to the training data. The distributions of the residuals are compared through their empirical cumulative density functions, both by  a visual inspection and by applying the Kolmogorov-Smirnov test \cite{Lopes2011}, as shown in Figure \ref{fig:cdfs}. The KS statistic measures the distance between any two CDFs. The null hypothesis of the test is that the two CDFs are equal. Setting $\alpha$, the probability of incorrectly rejecting the null hypothesis, induces a threshold for interpreting the KS statistic.  This confirms that it is likely that the train and test residuals follow the same distribution across the three data sets, implying that the model is not overfit.
	
	\subsection{Reconstruction of classic clear sky model}
	\begin{figure}[t!]
		\centering
		\includegraphics[width=7cm]{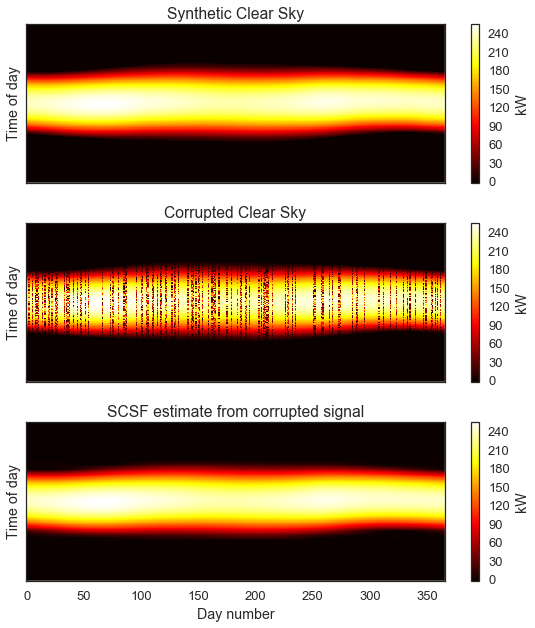}
		\caption{A synthetic clear sky PV signal from the \texttt{pvlib-python} pipeline (top) and corrupted with noise (middle). The SCSF estimate (bottom) recreates the original signal.}
		\label{fig:synthetic-cor-heatmap}
	\end{figure}
	As a final point of validation, we show that this method can exactly recreate the clear sky signal generated by a classic model pipeline, even after corrupting the signal with noise. We simulate one year of 5-minute data for a fixed-tilt system in Southern California using \texttt{pvlib-python} \cite{Holmgren2015}. We selected The Ineichin clear sky model, the Hay-Davies transposition model, and the Sandia Array Performance Model for the modeling pipeline. Generic but reasonable coefficients were selected for all models, as no particular effort was made to match this model to a specific real-world system. This model pipeline generates a clear sky irradiance signal, transposes it to the system array geometry, estimates the DC operating point of the PV system, and then applies a simple inverter efficiency model. Finally, we corrupt this synthetic clear sky signal by selecting $30\%$ of the days at random. Each value on the selected days is multiplied by a random number between $0$ and $1.1$. The resulting signal is then passed to the SCSF algorithm, and a new estimate of clear sky power is returned. As shown in Figure \ref{fig:synthetic-cor-heatmap}, the SCSF algorithm exactly reconstructs the uncorrupted signal, with an RMSE of $<0.5\%$ (relative to the magnitude of the original signal).
	
	\section{Implementation Considerations}
	The implementation of the SCSF algorithm used to generate the analysis in Section \ref{section-results} was written in Python and is available at \url{https://github.com/bmeyers/StatisticalClearSky}. Currently, the convex optimization subproblems are solved using \texttt{cvxpy} \cite{cvxpy} and \texttt{MOSEK} \cite{mosek}. \texttt{cvxpy} is a Python-based modeling language for convex optimization problems, and \texttt{MOSEK} is enterprise software for solving optimization problems. As implemented, the problem instances considered in this paper took on the order of minutes to solve on a 2.5 GHz Intel Core i7 processor with 16GB of RAM, which is reasonable given the off-line nature of this analysis. Still, this approach utilizes a very accurate but relatively slow interior-point method \cite{boyd2004}. This problem may be solved more efficiently through a direct implementation of the Alternating Direction Method of Multipliers (ADMM) \cite{Boyd2011}.
	
	\section{Future Work}
	Having developed the theory of the problem and proven the efficacy of the algorithm, the next step is to develop efficient open-source code for solving this class of problem. The authors will implement the ADMM approach for finding solutions to the subproblems, rather than relying on slower interior-point methods. In addition, the work will be published as an installable Python package, complete with documentation, examples, and a user guide.
	
	\section{Conclusion}
	We have presented an application of generalized low-rank modeling for solving the problem of estimating the clear sky signal from observed PV power data. This approach provides an efficient and model-agnostic method for automatically detecting clear periods in historical datasets, which is a marked improvement over previous approaches. Notably, this approach is resilient to system shading in the sense that the impact of site shading is part of the deterministic quasi-periodic baseline that the SCSF method selects for.  For multi-year data sets, the SCSF method automatically generates an estimate of system degradation at no extra cost as part of the data fitting process, without the need for irradiance data. In addition to automatic clear sky detection, the authors believe the signal produced by this algorithm has wide-ranging applications from unsupervised performance disaggregation to forecasting, due to its statistically unbiased fitting to observed data . 
	
	\section*{Acknowledgment}

	This work is supported by DOE/SU Contract DE-AC02-76SF00515. The authors would like to thank Prof. Stephen Boyd for his input and guidance.

	\ifCLASSOPTIONcaptionsoff
	\newpage
	\fi

	
	
	\bibliographystyle{IEEEtran}
	\bibliography{IEEEabrv,clearsky}
\end{document}